\documentstyle[aps,prb]{revtex}
\documentstyle[aps,prb]{multicol}
\newcommand{\beq}{\begin{eqnarray}}
\newcommand{\eeq}{\end{eqnarray}}
\begin{document}
\draft

\title
{ Positive Magneto-Resistance in Quasi-1D Conductors}
\author{Ivar Martin and Philip Phillips}
\vspace{.05in}

%
\address
{Loomis Laboratory of Physics\\
University of Illinois at Urbana-Champaign\\
1100 W.Green St., Urbana, IL, 61801-3080\\
\today}

%
\maketitle

\begin{abstract}

We present here a simple qualitative model that interpolates between the
high and low temperature properties of quasi-1D conductors.  At high temperatures
we argue that transport is governed by inelastic scattering whereas at low
temperatures the conductance decays exponentially with the electron dephasing length.
The crossover between these regimes occurs at the temperature at which
the elastic and inelastic scattering times become equal.  This model is shown
to be in quantitative agreement with the organic conductor $TTT_2I_{3-\delta}$.
Within this model, we also show that on the insulating side,
the positive magnetoresistance of the form
$(H/T)^2$ observed in $TTT_2I_{3-\delta}$ and other quasi-1D conductors
can be explained by the role spin-flip scattering plays in the electron
dephasing rate.  

\end{abstract}

\begin{multicols}{2}

\columnseprule 0pt
\narrowtext

At $T=0$, it is well-known that disorder precludes the existence of a metallic
state in strictly one and two-dimensional systems.\cite{local}
However, in the case of quasi-1D materials composed of coupled 1-d chains,
for example, conducting polymers and organic charge-transfer salts,
the situation is less clear. Abrikosov and
Ryzhkin\cite{abrikosov} argued that any coupling between a collection of
1-d chains will result in a non-zero conductivity at $T=0$.  However,
based on
a self-consistent diagrammatic approach, Prigodin and
Firsov\cite{prigodin,firsov}
showed that at $T=0$ a
quasi-1D system becomes insulating for sufficiently weak interchain
coupling.  Specifically, they showed that only when the interchain hopping matrix
element $t$ exceeds $0.3/\tau_{\rm el}$, does metallic transport obtain.
Here $\tau_{\rm el}$ is the elastic scattering time from static
impurities.  There are two major reasons why this conclusion is qualitatively
reasonable. 
First, if $t\ll 1/\tau_{\rm el}$, then an electron will scatter several times before
it hops onto another chain, thereby rendering  
interchain hopping ineffective against the localizing effect of disorder
on a single chain.  The second reason can be understood by considering
the dispersion relationship
\begin{equation}\label{disp}
\epsilon(p)= v_F(|p_z|-p_F)+t[\cos (ap_x/\hbar)+\cos(ap_y/\hbar)]
\end{equation}
for an anisotropic conductor, with $v_F$ and $p_F$ the Fermi 
velocity and momentum, respectively,
and $a$ the interchain spacing.  We assume here
that the chains form a square lattice in the $x-y$ plane 
(although real lattices are almost never perfectly square or rectangular,
this will prove inconsequential to our conclusions).  In the limit of  
strong elastic scattering, $1/\tau_{\rm el} \gg t$, broadening of the
Fermi surface 
exceeds the hopping part of the dispersion relationship, thereby
removing the 3-dimensional nature of the transport.  In this paper we are
concerned primarily with materials of this type.

In the opposite limit, when coupling between chains is relatively strong, 
$t > 1/\tau_{\rm el}$, the material behaves as an anisotropic 3D
metal at low temperatures.  
In this case, the effect of quasi-one-dimensionality can be completely absorbed 
into a diffusion constant that is anisotropic along the crystal 
axes.\cite{firsov,wolfle}  Such a material will remain conducting at the
lowest temperatures, unless a structural transition such as a
Peierls' distortion occurs.  

In this paper, we are concerned only with
materials that do not undergo
Peierls' transitions.  One of the outstanding problems
associated with the transport properties at low temperatures is the behavior
of the magneto-resistance on the insulating side.  For example,
in the charge transfer salt, $TTT_{2}I_{3-\delta}$,\cite{chakin} and
the conducting polymer p-phenylenevinylene (ppv),\cite{heeger} the 
magneto-resistance on the insulating side is positive and scales as
$(H/T)^2$, with $H$ the applied magnetic field and $T$ the temperature.  While
this functional form for the magneto-resistance is generally associated with
a competition between Zeeman and thermal energies, no formal account of this
phenomenon has been advanced within the context of strong or weak localization.
It is this problem we address in this paper.  Additionally, we focus on the origin of
the conductivity maximum exhibited by a wide class of quasi-1D
materials, most notably $TTT_{2}I_{3-\delta}$.   In $TTT_{2}I_{3-\delta}$,
the conductivity displays a maximum around $T_M=100$ K.  
Above $100$ K, the conductivity decreases as the temperature increases, indicating
the possible onset of a metallic state.  We first show that the conductivity
maximum represents a crossover from strong to weak localization.  At the crossover
the elastic and inelastic scattering times become equal.  Below $T_M$, we show
that within a strong localization account,
magnetic impurities can give rise to a positive magneto-resistance of the desired
form.  This effect arises from the field-dependence
of the spin-flip scattering contribution to the electron dephasing rate.  This 
explanation is in quantitative agreement with the observed experimental trends
for $TTT_{2}I_{3-\delta}$ as well as for (ppv).\cite{heeger}  The agreement with (ppv) suggests
that transport on the insulating side is dictated by strong localization
rather than by variable-range hopping as suggested previously.\cite{heeger}

We consider first the high temperature transport properties, that is,
$T>T_M$.  In most quasi-1D materials above $T_M$, the resistivity is of the form,
$\rho(T)=\rho_o+aT^\gamma$,
with $\gamma$ anywhere between 1 and 2.  The reason for the intermediate magnitude of
$\gamma$ compared to unity at high temperatures and 5 at low $T$ for regular
3D metals  is in the complex nature of the electron-phonon interaction in  
 quasi-1D materials.  This interaction is responsible for most of the temperature-dependent 
resistivity.\cite{chakin}  The contribution to the temperature-dependent resistivity from 
electron-electron interactions, on the other hand,  can oftentimes be neglected as a result of the low carrier density
 in  
most of the low-dimensional conductors.  Therefore, under the  assumption that elastic scattering 
from impurities and electron-phonon scattering are independent, the high-temperature
conductivity is given by the standard Drude formula 
\beq\label{hiT}
\sigma(T) &=& \frac{N n_1e^2\tau}{m}\\
\frac1\tau &=& \frac1\tau_{\rm el} + \frac1\tau_{\rm in} \approx \frac1\tau_{\rm el} + \frac1\tau_{\rm ph}
\nonumber
\eeq
 where $N$ is the number of chains per unit 
area, $n_1=2p_F/\pi\hbar$ is one-dimensional density of electrons, $m$ is the effective mass 
of the electrons, and $\tau_{ph}$ is the electron-phonon scattering time.

While  the low-temperature properties of  strongly coupled isotropic 
 materials can be understood  
in terms of
weak localization,\cite{magneto} the situation is quite
 different for ``real'' quasi-1D materials.
In such materials,
weak localization occurs only  when the inelastic scattering rate exceeds the elastic
scattering rate, $1/\tau_{\rm in} \gg 1/\tau_{\rm el}$. As a function of temperature then,
a crossover is expected from the transport being governed by 
localization (low temperatures) to a regime in which inelastic scattering dominates (high temperatures).
At low temperatures, transport is localized with a conductivity decaying
exponentially\cite{lee,thouless}
\beq\label{loT}
\sigma=\frac{e^2}{\hbar}N L_\phi e^{-\frac{L_\phi}{\xi}}
\eeq
with the Thouless length, $\L_\phi=v_F\sqrt{\tau_o\tau_{\phi}}$.
In Eq. (\ref{loT}), $\xi$ is the localization 
length, which in 1D-conductors is on the order of the elastic mean free path, 
$\ell=v_F\tau_{el}$, and  $\tau_{\phi}$ is the dephasing time,
which originates from all phase breaking processes.  This includes all inelastic processes and also
spin-flip scattering.
When the concentration of magnetic impurities is small,   
the dephasing rate is dominated over a wide temperature range by electron-phonon scattering. 

For a 1D system, the boundary separating the conductivity being given by the Drude result or the 
exponentially-localized Thouless form occurs when $1/\tau_{\rm in}\approx 1/\tau_{\rm el}$.    
Comparison of Eqs. \ref{hiT} and \ref{loT} reveals that at the temperature $T_M$, such that
$1/\tau_{\phi}(T_M) = 1/\tau_{\rm el}$, the high and low-temperature conductivities
predict a maximum value of
\beq\label{sigmamax}
\sigma_{\rm max} \simeq \frac{e^2}{\hbar}N v_F \tau_{\rm el}.
\eeq
Above and below $T_M$, the conductivities can be expressed in terms of 
$\sigma_{\rm max}$ and the temperature-dependent ratio $\tau_{\phi}(T)/\tau_{\rm el}$.  In particular,
if we assume that in the wide range of temperatures, which includes $T_M$, 
$1/\tau_{\phi}(T)\propto T^\gamma$, then
\beq\label{hilo}
\sigma(T)=\left\{\begin{array}{ll}
\sigma_{\rm max}(\frac{T_M}{T})^\frac\gamma2 e^{1-(T_M/T)^\frac\gamma2},\  &{\rm if} \quad T< T_M\\
\frac{2\sigma_{\rm max}}{1+(T_M/T)^\gamma},\ &{\rm if} \quad  T> T_M \end{array}\right.
\eeq

Despite the simple considerations that led to  Eq. (\ref{hilo}), this expression
is of extreme utility in 
understanding the nature of the maximum in the conductivity.  The exponent $\gamma$ can be easily 
extracted form the experimental data.  If the ratio of the conductivity maximum 
 to the room temperature conductivity is 
$\sigma_{\rm max}/\sigma_{\rm RT}\simeq (T_M/T)^\gamma/2$, then it is likely that the observed 
transition is of the type described here.
In fact, this condition is satisfied for the ($TTT_{2}I_{3-\delta}$) samples studied by Khanna,
 et. al.\cite {chakin}  In their  samples the conductivity maximum occurred at
$T_M\approx 100\ {\rm K}$ and the thermal exponent $\gamma$ is approximately 2.\cite{somoano}  Within our 
picture this yields $\sigma_{\rm max}/\sigma_{\rm RT}\sim 4$, which is reasonably close to the
observed ratio of $2-3$.  At low temperatures, the conductivity declines rapidly which is
consistent with the first of the expressions in Eq. (\ref{hilo}).
However, at very low temperatures, the deviation from the predicted behavior becomes significant.
For example, at $T=4 \ {\rm K}$ conductivity is roughly 1 to 10 percent (depending on the degree of
 disorder) of $\sigma_{\rm RT}$.
Although this value is small, it is significantly larger than that predicted from the first of
the expressions in Eq. 
(\ref{hilo}). The source of this discrepancy is straightforward to pinpoint.
  In obtaining Eq. (\ref{hilo}) we used that $1/\tau_{in}(T)\propto T^\gamma$, which is 
essentially an extrapolation of the high temperature electron-phonon scattering rate to low 
temperatures, which can be expected to break down, for instance, when the thermal phonon wavelength,
$\lambda_{ph}$, becomes comparable to the mean free path (dirty limit).
Even more importantly, up to this point, we have neglected all other dephasing mechanisms besides 
electron-phonon scattering.  As the temperature decreases, the relative role of the electron-electron
interaction increases.  Also, spin-flip scattering has not been incorporated. 
In fact, transport measurements reveal the presence of a positive
magnetoresistance of the form $(H/T)^2$ below 20 K.\cite{chakin} 
We now show that inclusion of spin-flip scattering is essential to explaining the positive
magnetoresistance.  Hence, this effect influences the conductivity as well. 

It is well known that the spin-flip scattering suppresses localization.\cite{altshuler}
Unlike electron-phonon scattering, spin-flip scattering is essentially temperature-independent in the absence
of a magnetic field for temperatures higher than the Kondo temperature, $T_K$.\cite{kondo}
Below $T_K$ the spin-flip scattering rate gradually decreases and disappears completely at $T=0$ for
spin $S=1/2$ impurities.
When  spin-flip scattering is no longer negligible compared with the other (inelastic) dephasing 
processes, we must explicitly include it in the dephasing rate:
\beq
\frac1{\tau_{\phi}}=\frac1{\tau_{\rm in}}+\frac1{\tau_{s}}
\eeq
where $\tau_s$ is the spin-flip scattering time and $\tau_{\rm in}$ now represents all
inelastic processes, such as electron-electron interactions and  electron-phonon scattering.

Of these processes, the only one that couples to a magnetic field is spin-flip scattering. 
In a non-magnetic 1D system, the fact that flux cannot be enclosed
signals a vanishing of the localization correction to the magneto-resistance. 
Hence, magnetic
impurities offer a channel to circumvent the apparent lack of magnetic coupling in
1D systems.
In the presence of a field, the spin-flip scattering rate decreases because the
spins partially align with the field. As a consequence, $L_\phi$ grows and 
the conductivity given by the first of the Eqs. (\ref{hilo}) decreases.  Hence, the magnetoresistance
is positive. To determine the functional form of the magnetoresistance, we calculate
the spin-flip scattering time in the presence of the field.  
To second order in the exchange interaction, the spin-flip scattering
rate is given by\cite{diskondo}
\beq
\frac1{\tau_s}=\frac1{\tau_s^o}
\left(\frac12+\frac{2\beta h}{\sinh(2\beta h)}\right)
\eeq
where $\tau_s^o$ is the spin-flip scattering time in the absence of magnetic field, $\beta=1/(k_B T)$ 
and $h=g\mu_B H/2$.  Here $\mu_B$ is the Bohr magneton and $H$ is the applied magnetic field.  
In the experimental range of interest $\beta h< 1$.
Consequently, we expand $\sinh x$ to obtain, 
\begin{equation}
\frac1{\tau_s}=\frac1{\tau_s^o}\left(\frac32-\frac23(\beta h)^2\right)
\end{equation}
Hence, the spin-flip scattering rate decreases as $(H/T)^2$. If we simply insert this
expression into the dephasing length, we find immediately that in the leading
power of $H/T$, the magnetoconductance is
\beq\label{sigmah}
\frac{\sigma(H)-\sigma(0)}{\sigma(0)}=-\sqrt{\frac{\tau_\phi^3}{\tau_{el}\tau_s^{o2}}}\left(\frac{\mu_BH}
{3k_BT}\right)^2
\eeq
In general, $\tau_\phi$ is a function of temperature since all inelastic processes are 
temperature-dependent.  But for low enough temperature (but still higher 
than $T_K$!),
spin-flip scattering can become the dominant dephasing mechanism, and then from 
$\tau_\phi\approx \tau_s$ and equation (\ref{sigmah}) it immediately follows that the
magnetoconductance 
\beq\label{sigmahs}
\frac{\sigma(H)-\sigma(0)}{\sigma(0)}=-\sqrt{\frac{\tau_s^o}{\tau_{el}}}\left(\frac{\mu_BH}{3k_BT}\right)^2
\eeq
is of the $(H/T)^2$ form often seen experimentally.
In particular, in the experiment of Khanna and coworkers\cite{chakin} such a
 positive magnetoresistance
was observed below 20 K.  No anisotropy was observed, which strongly supports the spin nature of the
magnetoresistance.  
If we assume that the dominant dephasing mechanism is given by 
electron-phonon scattering, we find that
$\tau_{\phi} \approx \tau_{\rm el}(T/T_M)^\gamma$.  Form this we estimate that
 the spin-flip scattering 
time $\tau_s\approx \tau_\phi(20K)\approx 25 \tau_{\rm el}$.  Applying Eq. (\ref{sigmahs}) then gives 
$\Delta \sigma(H)/\sigma(0)\simeq 0.2(H/T)^2$, where $H$ is in Tesla and $T$ in Kelvin.
The corresponding prefactor extracted  from the experimental data\cite{chakin} is between 0.1 and 0.2
which is in excellent agreement with our qualitative prediction.
At even lower temperatures (below 4 K) the field and temperature dependence of the magnetoresistance become
weaker.  At the same time, the zero-field conductivity continues to decrease.  These observations are 
consistent with the quenching of the impurity spin by the conduction electrons below the Kondo temperature, 
$T_K$.

We have shown quite generally that quasi-1D systems in which localization
effects are dominant acquire a positive magneto-resistance of the form 
$(H/T)^2$ through spin-flip scattering from magnetic impurities.
This correction was shown to be in quantitative agreement with the experiments
on the charge transfer salt $TTTT_2I_{3-\delta}$. A similar positive magneto-resistance
was observed in the conducting polymer ppv.\cite{heeger} Heeger
and co-workers\cite{heeger} found that neither the Mott variable-range
hopping model\cite{mott} nor the Efros-Shklovskii Coulomb gap
model\cite{es} could account for the $(H/T)^2$ trend in the 
magneto-resistance.  The model presented here offers a natural explanation
of this experimental trend.  Experimental susceptibility measurements on
ppv could be used to test the presence of magnetic impurities and hence confirm
the scenario presented for the origin of the positive magneto-resistance.

\acknowledgments
This work was supported by the NSF grant No. DMR94-96134
and the donors of the Petroleum Research Fund administered by the American
Chemical Socitey.

\end{multicols}
\end{document}